\documentclass[a4paper,11pt]{article}
\usepackage{pos}
\usepackage{lineno}
\usepackage[subrefformat=parens,labelformat=parens]{subfig}
\usepackage{xspace}
\newcommand*{\TeV}{\ifmmode {\mathrm{\ Te\kern -0.1em V}\xspace}\else
                  \textrm{Te\kern -0.1em V}\xspace\fi}%
\newcommand*{\GeV}{\ifmmode {\mathrm{\ Ge\kern -0.1em V}\xspace}\else
                  \textrm{Ge\kern -0.1em V}\xspace\fi}%
\newcommand*{\MeV}{\ifmmode {\mathrm{\ Me\kern -0.1em V}\xspace}\else
                  \textrm{Me\kern -0.1em V}\xspace\fi}%

\let\tev=\TeV
\let\gev=\GeV

\def\st{sim\_telarray\xspace}
\def\lat{\textit{Fermi}-LAT\xspace}

\def\wide{M6D1a\xspace}

\title{The Cherenkov Telescope Array: layout, design and performance}
 \ShortTitle{CTA: layout, design and performance}

\author*[a]{Orel Gueta}

\affiliation[a]{Deutsches Elektronen-Synchrotron (DESY),\\
  Platanenallee 6, Zeuthen, Germany}

\forColl{CTA Consortium and the CTA Observatory} 

\emailAdd{orel.gueta@desy.de}

\abstract{The Cherenkov Telescope Array (CTA) will be the next generation very-high-energy gamma-ray observatory. CTA is expected to provide substantial improvement in accuracy and sensitivity with respect to existing instruments thanks to a tenfold increase in the number of telescopes and their state-of-the-art design. Detailed Monte Carlo simulations are used to further optimise the number of telescopes and the array layout, and to estimate the observatory performance using updated models of the selected telescope designs. These studies are presented in this contribution for the two CTA stations located on the island of La Palma (Spain) and near Paranal (Chile) and for different operation and observation conditions.}

\FullConference{37$^{\rm{th}}$ International Cosmic Ray Conference (ICRC 2021)\\
		July 12th -- 23rd, 2021\\
		Online -- Berlin, Germany}

\begin{document}
\maketitle

\section{Introduction}

The Cherenkov Telescope Array (CTA)~\cite{CTA} will be the next generation very-high-energy (VHE; $10 \GeV \lesssim E \lesssim 100 \TeV$) gamma-ray observatory. 
CTA will be composed of more than 50 imaging atmospheric Cherenkov telescopes (IACTs) and will be built at two sites located on the island of La Palma (Spain; 17.89W, 28.76N; 2158~$m$ altitude) and near Paranal (Chile; 70.32W, 24.68S; 2147~$m$ altitude). 
IACTs detect gamma rays by recording the Cherenkov light induced by extensive air shower particles produced in the interaction of the gamma ray with particles in the atmosphere~\cite{1989ApJ...342..379W}.
CTA will be sensitive to gamma rays in a wide energy range, from 20 \gev to beyond 300 \tev, thanks to the use of three types of telescopes of varying sizes; namely the Large-Sized Telescopes (LSTs), Medium-Sized Telescopes (MSTs) and Small-Sized Telescopes (SSTs).
Its sensitivity is expected to be about a factor of five to ten better than current IACTs across the entire energy range, providing the capability to e.g., perform deep surveys of various sky regions. 
The short-term sensitivity of CTA will be a few orders of magnitude better than that of \lat. 
This would enable the measurement of very-fast variability in active galactic nuclei flares and increase the likelihood of detecting short-timescale transient phenomena like gamma-ray bursts or black-hole mergers.
CTA is expected to provide a substantial improvement in angular and energy resolution. The angular resolution of CTA is expected to reach around one arc minute at high energies. This, together with the large field of view ($4.5^\circ$ -- $8.5^\circ$), will enable detailed imaging of extended gamma-ray sources. The improved energy resolution, reaching around 5\% at 1~\tev, will make it easier to measure spectral cutoffs and detect spectral features such as those expected from dark matter.

The telescope simulation models used in previous CTA performance studies~\cite{Maier:2019afm} were updated and new estimates were derived.
These estimates, together with the telescope layout optimisation process, are presented in this contribution.
The number of telescopes considered here is not the final one and is subject to change. 

\section{Simulation and analysis}

To optimise the CTA telescope layout and derive its performance,
detailed Monte Carlo simulations were generated.
The process starts with simulations of extensive air showers and the Cherenkov light they induce using the CORSIKA program (version 7.7100)~\cite{corsika}.
The Cherenkov photons from the air shower are propagated through the telescopes using the \st simulation package (version 2020-06-28)~\cite{Simtelarray}.
Detailed simulation models of the CTA telescopes, are included in \st.
These models were updated for the purpose of this study. The update process included:
\begin{itemize}
    \item obtaining an updated atmospheric model for La Palma and geomagnetic field values for both sites;
    \item performing detailed ray-tracing simulations of optical elements to e.g., update the reference model used in \st to model the shadowing on the camera as a function of off-axis angle due to various telescope components.
    \item collecting lab and on-site prototype measurements of various telescope components simulated in \st;
    \item tuning simulation parameters to fit the measurements where necessary;
    \item deriving appropriate trigger threshold levels;
    \item estimating the expected night-sky background light level in each pixel.
\end{itemize}

A summary of the samples simulated for each site, including the number of events per particle type per site and the energy ranges, is given in Table~\ref{tab:simulatedParticles}. Three zenith angles (20$^\circ$, 40$^\circ$ and 60$^\circ$) and both north and south pointings were simulated, with all telescopes pointing parallel to each other.
\begin{table}[htbp]
\centering
\begin{tabular}{c|c|c}
\hline
Particle type & Energy range [TeV] & Number of events \\ \hline
Point source gamma & 0.003 -- 330 & $1\mathrm{e}{9}$ \\ 
Diffuse gamma & 0.003 -- 330 & $5\mathrm{e}{9}$ \\ 
Electron/Positron & 0.003 -- 330 & $1\mathrm{e}{9}$ \\
Proton & 0.006 -- 600 & $20\mathrm{e}{9}$ \\ \hline
\end{tabular}
\caption{\label{tab:simulatedParticles} Number of events simulated per particle type and the corresponding energy ranges. The same number of events was simulated for both the northern and southern sites.}
\end{table}

\begin{figure}
\centering\includegraphics[trim=5mm 5mm 20mm 20mm,clip, width=0.48\linewidth, page=1]{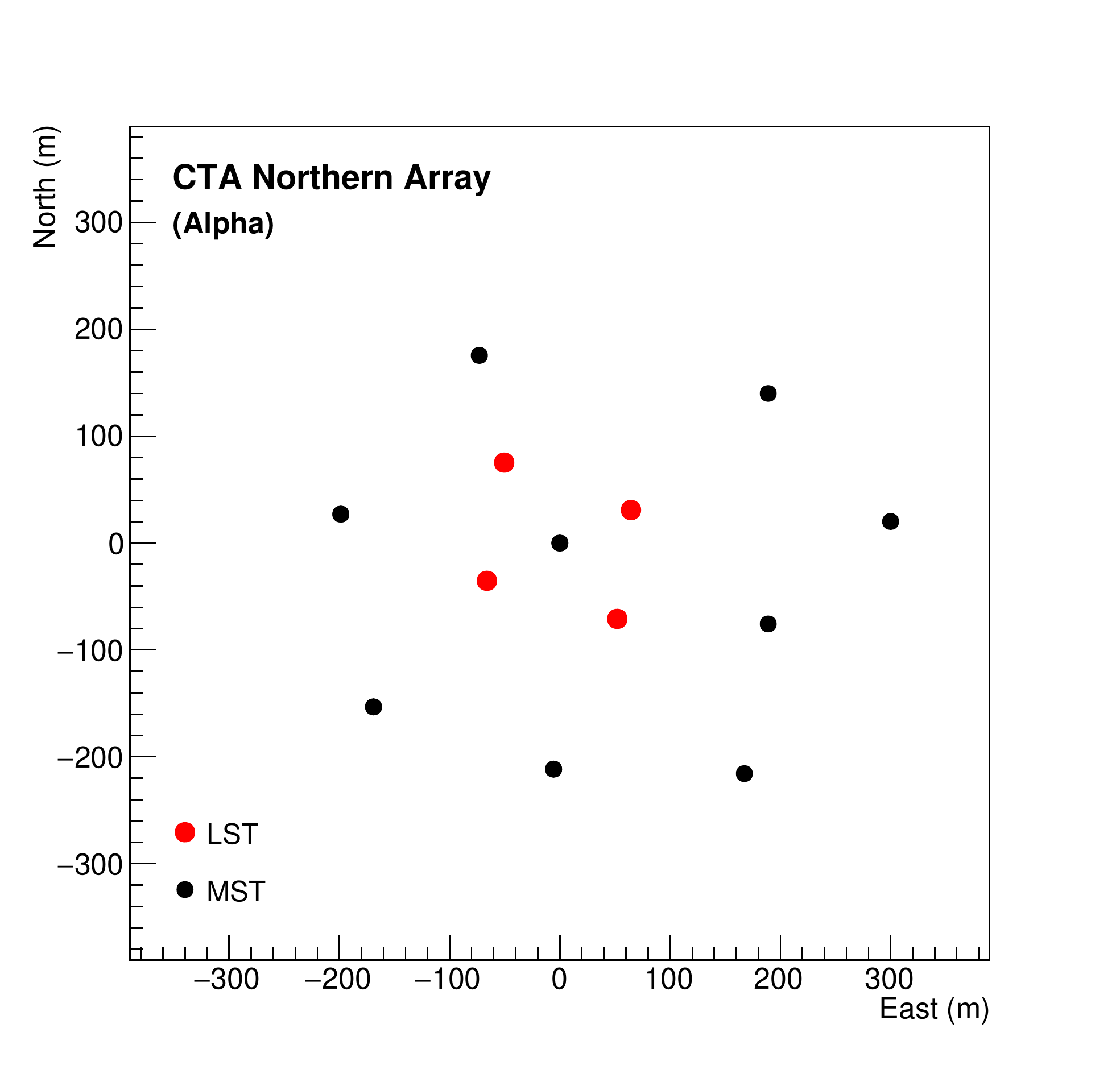}
\centering\includegraphics[trim=5mm 5mm 20mm 20mm,clip, width=0.48\linewidth, page=1]{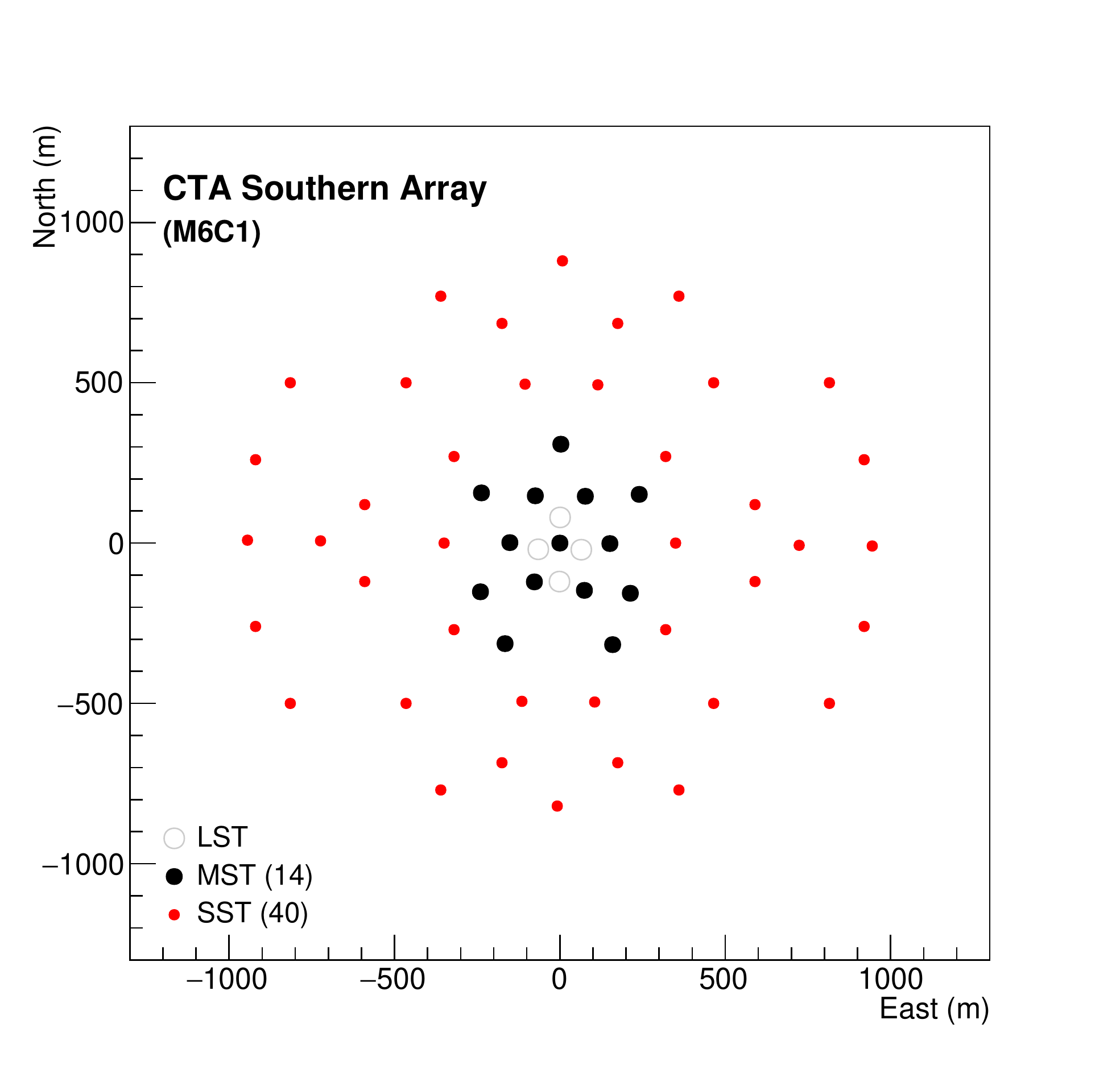}\\
\centering\includegraphics[trim=5mm 5mm 20mm 20mm,clip, width=0.48\linewidth, page=1]{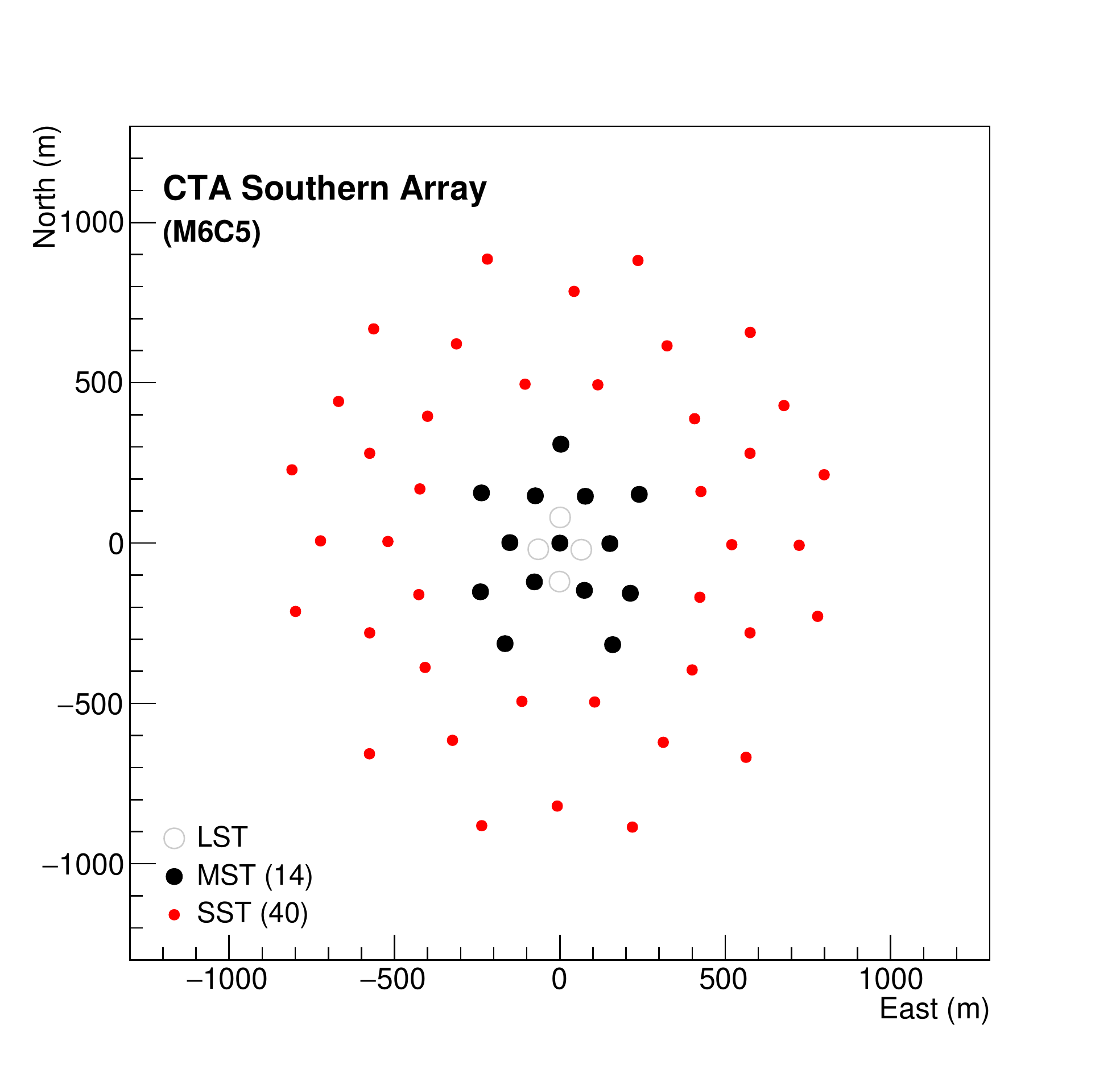}
\centering\includegraphics[trim=5mm 5mm 20mm 20mm,clip, width=0.48\linewidth, page=1]{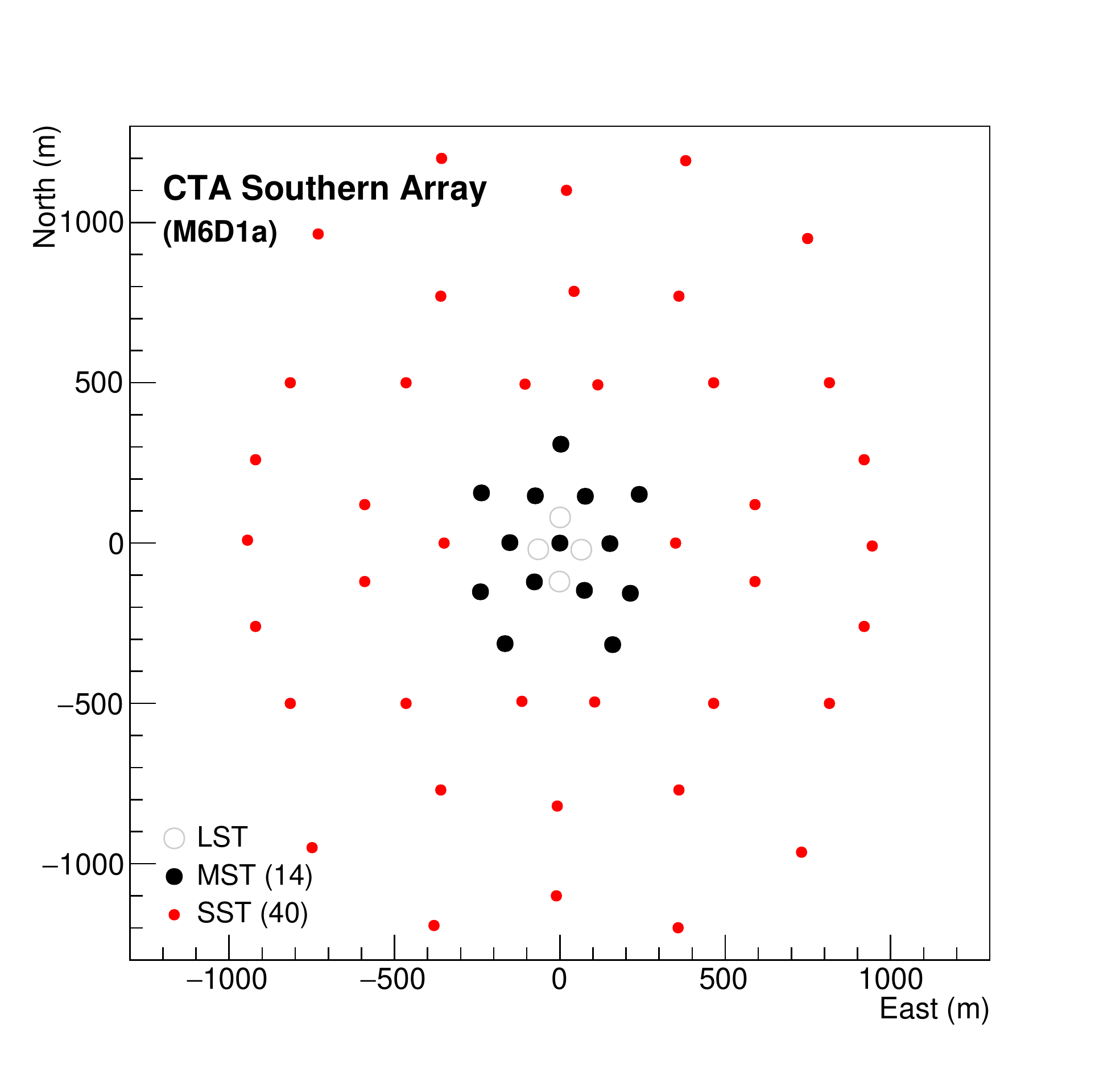}

\caption{\label{fig:arrays}
Telescope layouts for the CTA northern La Palma site (top left) and the southern Paranal site (top right, bottom).
Telescope types are indicated in the legend.
The Alpha layout shown for the northern site will be built during the construction phase.
Three possible layouts are shown for the southern site, where the difference is in the layout of the SSTs.
The LSTs shown in the southern layouts were not included in the simulations. They will potentially be added to the array in the future.
Note that the area covered by the southern layout is approximately 19 times that of the northern layout.}
\end{figure}

The telescope layouts for a larger number of telescopes on both sites were extensively optimised in a previous study~\cite{CTAarrays}. 
In order to further optimise the layout with fewer telescopes as considered here, a total of 180 telescope positions were simulated in the southern site and 84 positions in the northern site.
Out of those simulated positions, many telescope configurations were considered.
Figure~\ref{fig:arrays} shows one configuration for the northern site, the Alpha configuration to be built in the construction phase, and three configurations for the southern site. 
The three telescope layouts in the southern site shown in Figure~\ref{fig:arrays} are the leading candidates to be selected for construction. A final decision on the layout is dependent on the number of telescopes to be built. Also shown are the positions of four LSTs not included in the simulations. Those  will potentially be added to the array in the future.

The reconstruction and analysis of the simulation output were performed using the Eventdisplay~\cite{Eventdisplay} analysis software (version prod5\_d20200702).\footnote{Dedicated CTA reconstruction software is currently under development and is expected to provide improved performance compared to the one presented here.} The main steps of the analysis consist of waveform integration, image cleaning, stereoscopic reconstruction and cut optimisation. A detailed description of the reconstruction can be found elsewhere (see e.g., Ref.~\cite{2013APh....43..171B}).

All of the performance evaluations shown in Figures~\ref{fig:diffsens}~--~\ref{fig:shortSens} are for point-like gamma-ray sources observed at a zenith angle
of 20 deg, located either at the centre of the field of view (on-axis) or 3$^\circ$ -- 4$^\circ$ off axis as indicated in the figures. Results were averaged between the two north and south pointings.
Further zenith angles, off-axis angles and night-sky background levels were taken into account in the layout optimisation, but are not shown here for reasons of brevity.

\section{Layout optimisation and performance}
\label{sec:performance}

The main criteria typically used to evaluate the performance of IACTs are effective collection area, angular resolution, energy resolution, residual background rate and differential sensitivity.
All were considered in the layout optimisation, but only the effective collection area, angular resolution and differential sensitivity are discussed here.

\begin{figure}[ht!]
\centering\includegraphics[trim=5mm 0mm 18mm 10mm,clip, width=0.49\linewidth, page=1]{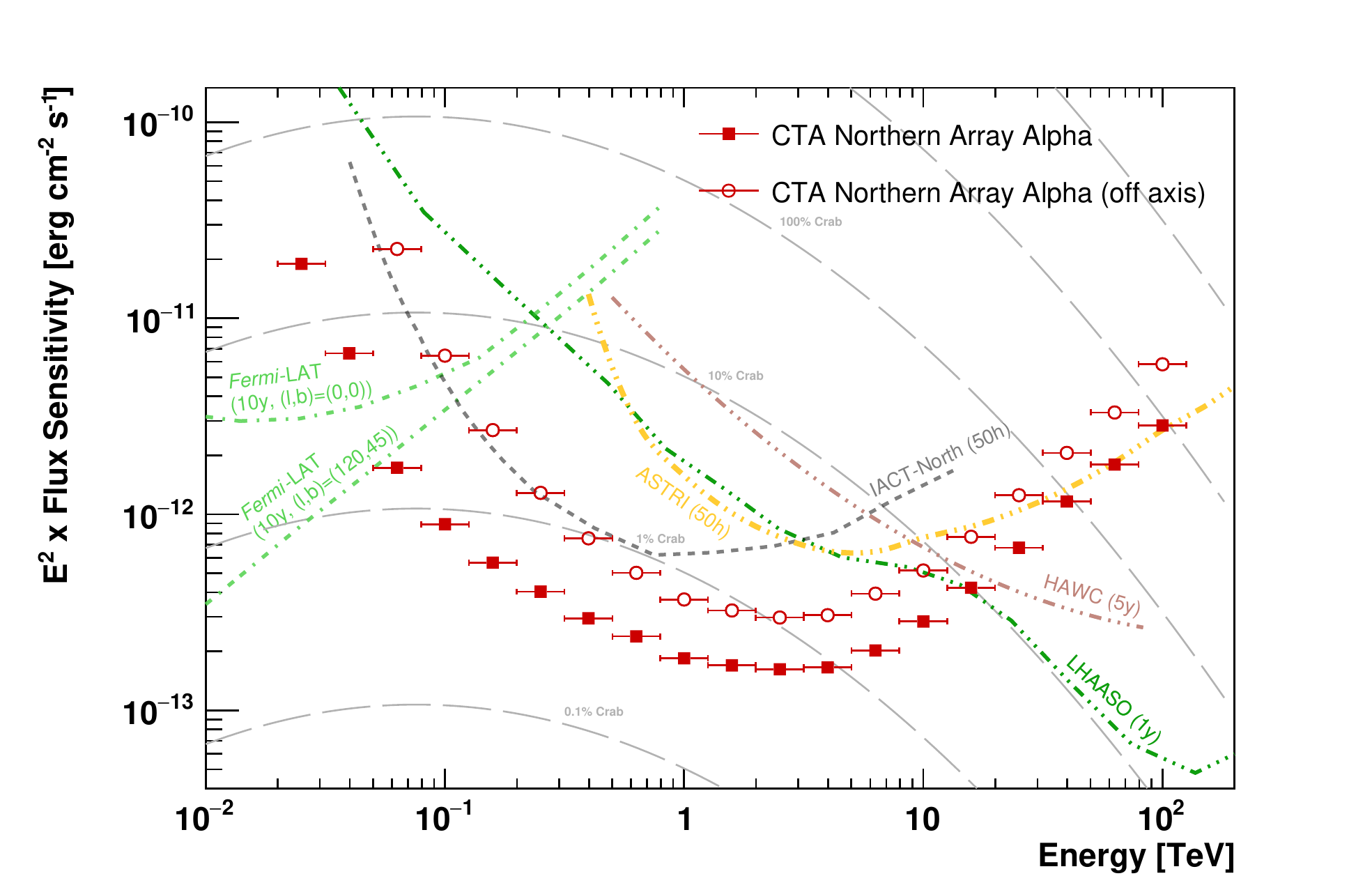}
\centering\includegraphics[trim=5mm 0mm 18mm 10mm,clip, width=0.49\linewidth, page=1]{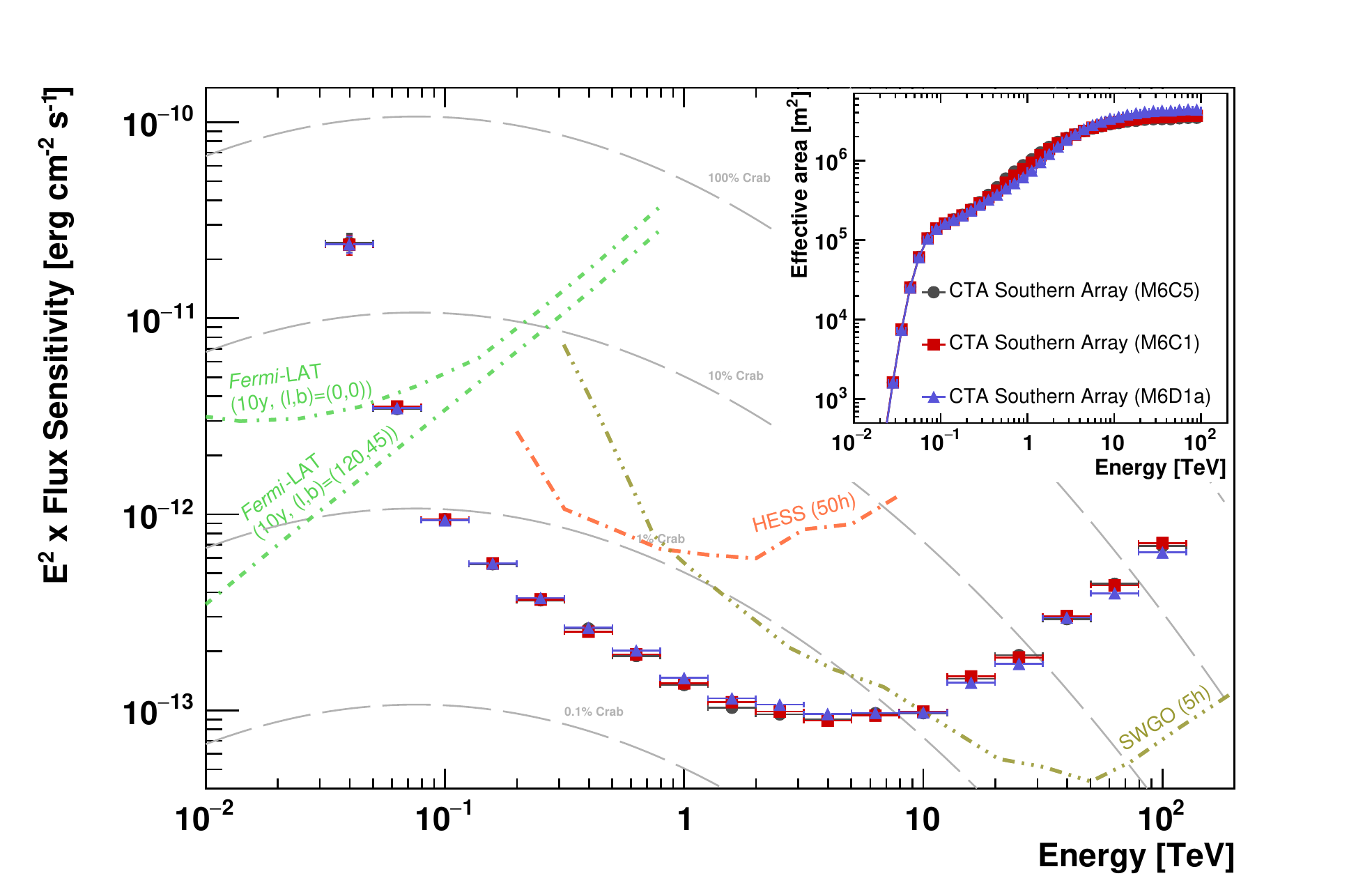}
\caption{\label{fig:diffsens}
The CTA differential point-source sensitivity as a function of reconstructed energy in the northern array site (left) and for the various telescope layouts in the southern array site (right), assuming 50 hour observation time, pointing to 20 degrees zenith and averaged between pointing north and south.
Filled markers represent observations assuming a gamma-ray source located at the centre of the field of view, while empty markers are for a source located 3$^\circ$ -- 4$^\circ$ off axis (northern site only).
The differential sensitivities of other instruments~\cite{Latperf, HESSperf, MAGICperf, VERITASperf, ASTRIPerf, HAWCperf, LHAASOsens, Albert:2019afb} are shown for comparison.
The IACT North curve represents the best-of sensitivity of MAGIC and VERITAS.
The curves for \lat and HAWC are scaled by a factor 1.2 relative those provided in the references, to account for the different energy binning.
A comparison between the effective collection areas of the different telescope layouts in the south, calculated assuming 30 minutes observation time, is shown in the inset in the figure.}
\end{figure}

The differential sensitivity, defined as the minimal flux required to detect a point-like source ($\sigma > 5$), is the primary metric used to decide between telescope layouts.
To calculate the sensitivity, apart from the detection requirement, at least ten detected gamma rays and a minimal signal to background ratio of 1/20 in each energy bin were required.
The analysis cuts in each energy bin were optimised for best flux sensitivity in the northern site.
In the southern site both flux sensitivity and angular resolution were taken into account in the optimisation process, with a higher weight given to the angular resolution.
Figure~\ref{fig:diffsens} shows the on-axis sensitivity achieved with the various telescope layouts on both sites for an assumed 50 hour observation time (the optimal analysis cuts depend on the observation time).
For the northern site, also the off-axis sensitivity is shown, where such off-axis observations at 3$^\circ$ -- 4$^\circ$ from the camera centre will be possible for the first time thanks to the large field of view of CTA cameras.
The differential sensitivities of other instruments~\cite{Latperf, HESSperf, MAGICperf, VERITASperf, ASTRIPerf, HAWCperf, LHAASOsens, Albert:2019afb} are shown for comparison.
These should provide only a rough comparison of the sensitivity of the different instruments, as the method of calculation and the criteria applied are not identical.
The MAGIC and VERITAS sensitivities were combined by taking the more sensitive value of the two arrays in each bin and are represented by the IACT North curve.

A comparison between the effective collection areas of the three candidate layouts of the southern site are shown in the inset in Figure~\ref{fig:diffsens}.
The effective collection area is defined as the differential gamma-ray detection rate divided by the differential flux of incident gamma rays.
The effective collection area was calculated assuming 30 minutes observation time and optimising the cuts for best sensitivity.

CTA will outperform current generation IACTs by about a factor five across the entire energy range. The energy coverage will expand as well, in particular at high energies, reaching hundreds of \tev, compared to $\sim 10 \tev$ with current IACTs.
Above 10 -- 20 \tev, HAWC, LHAASO and SWGO are more sensitive than CTA, albeit with a worse angular resolution (at 100 \tev, the angular resolution of HAWC and SWGO is about 0.1$^\circ$ and the LHAASO one is around 0.2$^\circ$~\cite{HAWCperf, Aharonian:2020iou, Albert:2019afb}).

In the southern site, a difference of about 10 -- 20\% in sensitivity is seen between the layouts. In particular, the \wide layout, where the SSTs are spread further apart (see the bottom right of Figure~\ref{fig:arrays}), is less sensitive around 5 \tev and more sensitive above 10 \tev. 
This can be explained by the differences seen between the layouts in the effective collection area.

\begin{figure}[ht!]
\centering\includegraphics[width=0.49\linewidth]{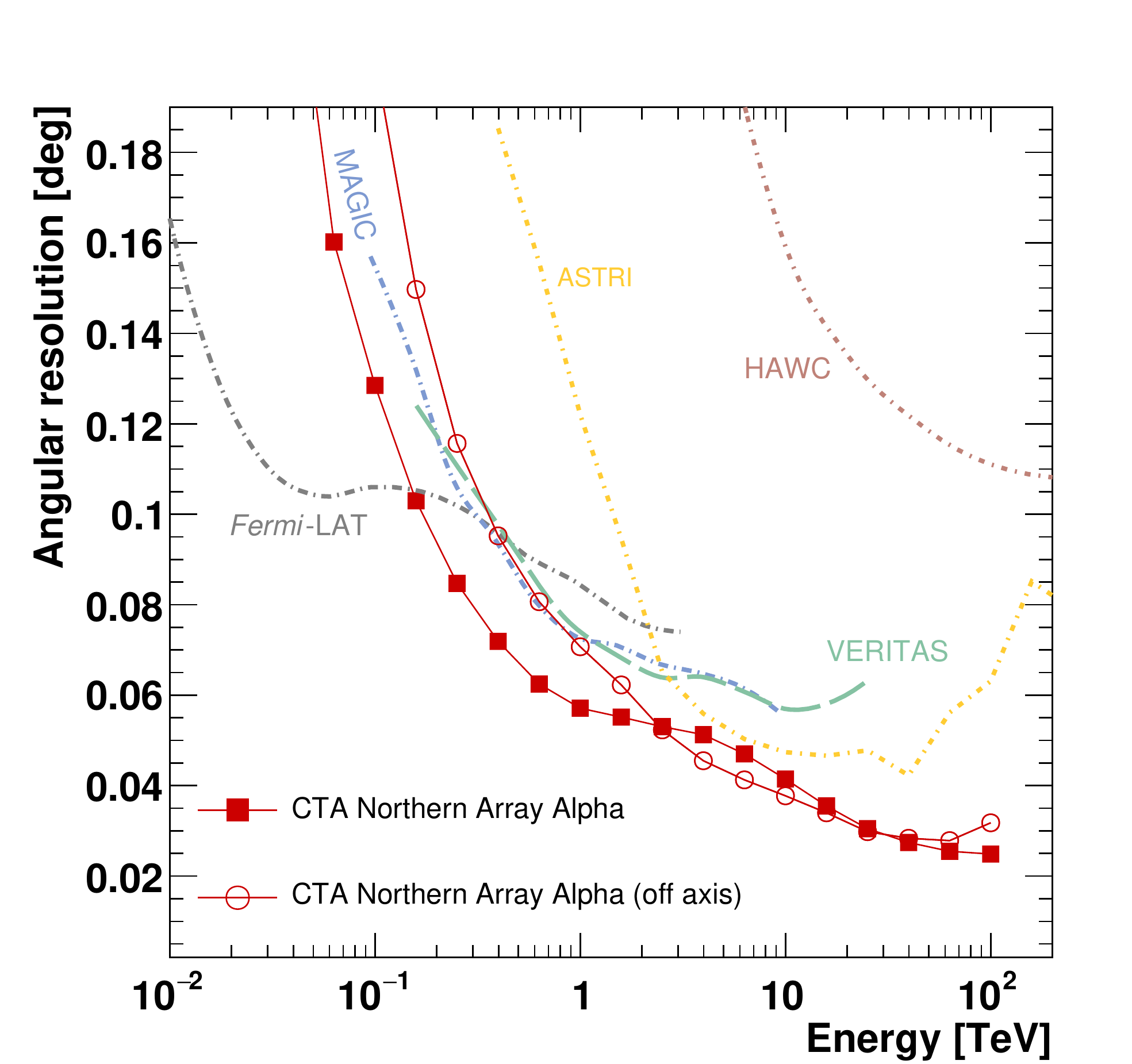}
\centering\includegraphics[width=0.49\linewidth]{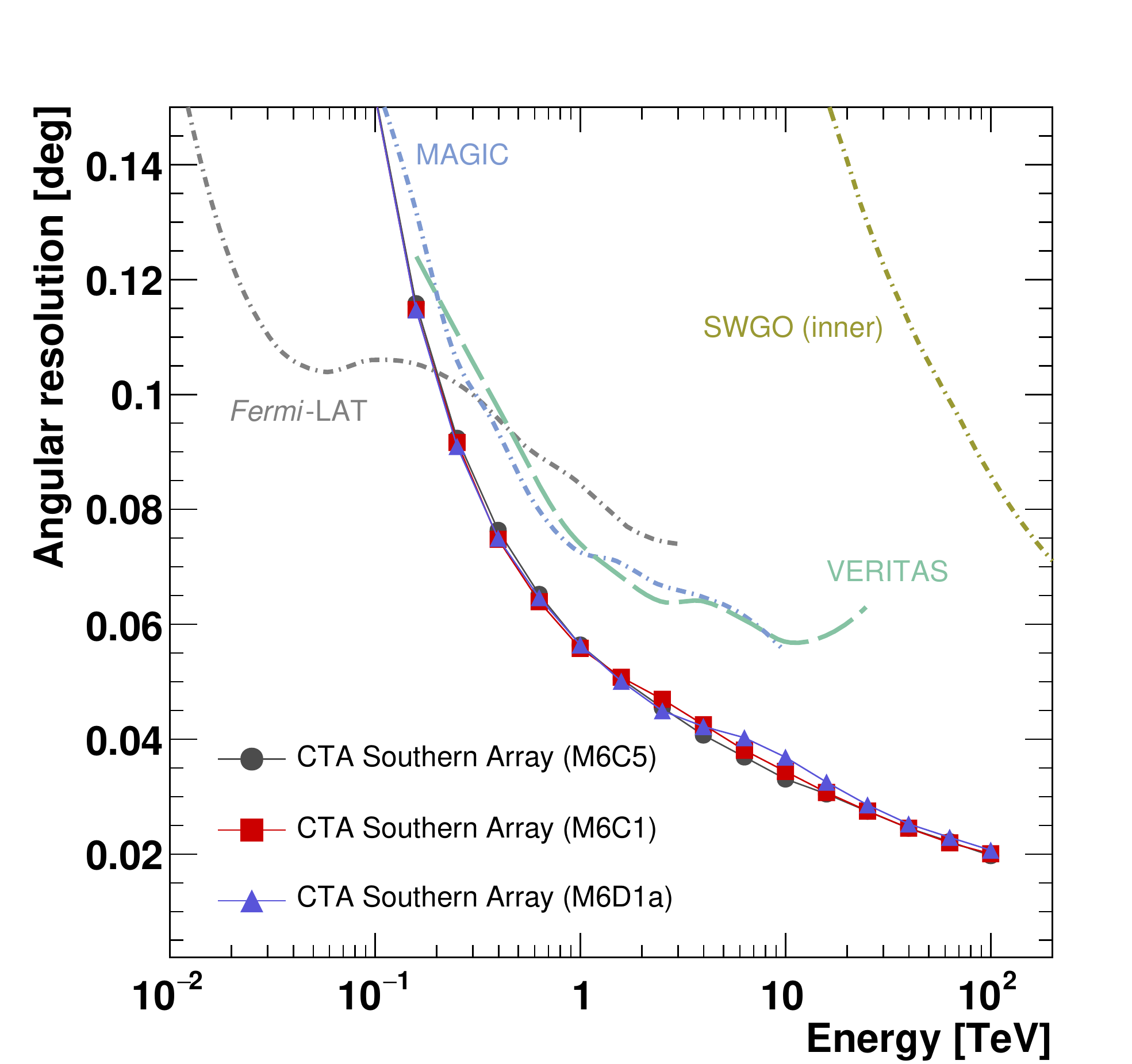}
\caption{\label{fig:angRes}
Angular resolution as a function of reconstructed energy for the northern (left) and southern site (right) of CTA.
The angular resolution curves of the various layouts are indicated in the legend, where filled markers represent observations assuming a gamma-ray source located at the centre of the field of view, while empty markers are for a source located 3$^\circ$ -- 4$^\circ$ off axis (northern site only).
The angular resolution of other instruments is shown for comparison~\cite{Latperf, MAGICperf, VERITASperf, ASTRIPerf, HAWCperf, Aharonian:2020iou, Albert:2019afb}.}
\end{figure}

The angular resolution as a function of the reconstructed energy obtained with the various telescope layouts on both sites is shown in Figure~\ref{fig:angRes}.
The angular resolution in each energy bin is defined as the angle containing 68\% of the reconstructed gamma-ray events relative to the simulated gamma-ray direction.
A small improvement of less than 0.01$^\circ$ in angular resolution is observed around 10 \tev between the layouts in the southern site, where the angular resolution of the \wide layout is worse than of the other layouts.
The angular resolution of other instruments~\cite{Latperf, MAGICperf, VERITASperf, ASTRIPerf, HAWCperf, Aharonian:2020iou, Albert:2019afb} is shown in Figure~\ref{fig:angRes} for comparison.
CTA will provide a significant improvement in angular resolution compared to other instruments, ranging between 0.02$^\circ$ to 0.2$^\circ$. It should be noted that the analysis performed here was only partially optimised for best angular resolution.
Better angular resolution is obtainable with appropriately optimised cuts in case of e.g., morphology studies of bright sources.

\begin{figure}[ht!]
\centering\includegraphics[trim=7mm 3mm 19mm 5mm,clip,width=0.65\linewidth]{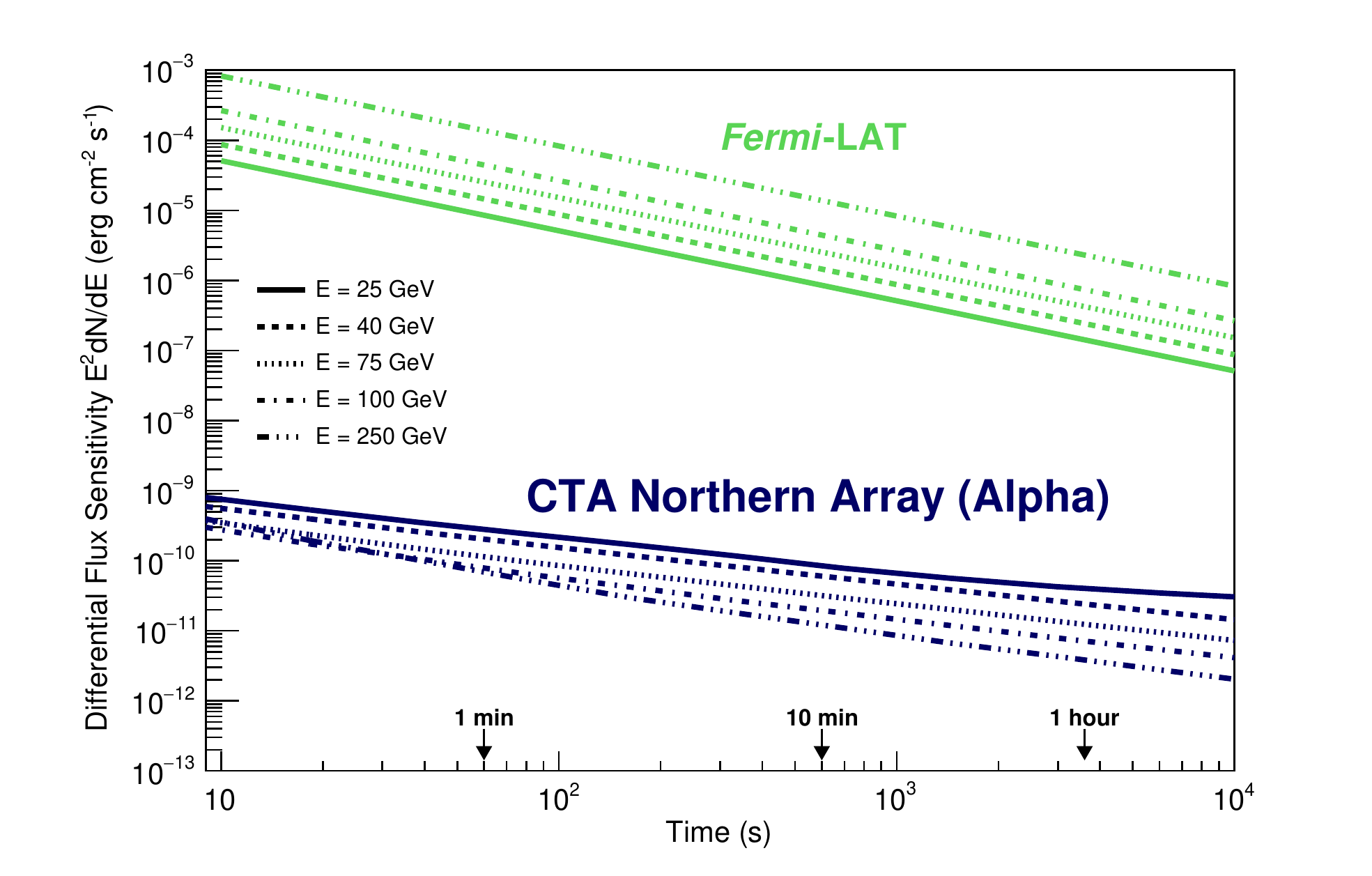}
\caption{\label{fig:shortSens}
Differential flux sensitivity of the northern site of CTA at selected energies as function of observing time for 20 degrees zenith observation and averaged between pointing north and south, in comparison with \lat (Pass 8 analysis, extragalactic background, standard survey observing mode). 
}
\end{figure}

The short-transient phenomena discovery potential of the northern site of CTA is demonstrated in Figure~\ref{fig:shortSens}, where the integral sensitivity in each energy bin for selected energies is shown as a function of observation time.
The short-term sensitivity of \lat is shown for comparison.
CTA is a few orders of magnitude more sensitive than \lat for such short observation times.
The discovery potential of CTA for e.g., gamma-ray bursts, is therefore significantly higher than that of \lat. However, that is true only for sources which are in the field of view. The \lat field of view is substantially larger at 2.4 sr, making such occurrences much more likely.
To enhance the efficiency of CTA in discovering short-term phenomena, CTA telescopes were designed with fast repointing capabilities. The LSTs will be able to repoint within 20 seconds of receiving an alert of a potential transient source.

\section{Conclusions}


The performance of various CTA telescope layouts was estimated and compared.
The northern site of CTA is expected to be about five times more sensitive than MAGIC and VERITAS, while the southern site is expected to be an order of magnitude more sensitive than HESS.
At energies above 10 -- 30 \tev, CTA is less sensitive than water Cherenkov detector arrays, but provides significantly better angular and energy resolution.
The differences between the sensitivities of the layouts in the southern site are of the order of 20\% at energies above 1 \tev.
The short-term sensitivity of CTA is expected to be a few orders of magnitude better than \lat, making it a great instrument for short-term phenomena detections, in particular for serendipitous discoveries of sources in the field of view.

\acknowledgments

\noindent This work was conducted in the context of the CTA Consortium and CTA Observatory.

\noindent We gratefully acknowledge financial support from the agencies and organizations listed here: 
\url{http://www.cta-observatory.org/consortium_acknowledgments}.
We also would like to thank the computing centres that provided resources for the generation of the instrument response functions, see Ref.~\cite{CTAPerf} for a full list of contributing institutions.

\clearpage
\section*{Full Authors List: the CTA Consortium and the CTA Observatory}
%
%
\scriptsize
\noindent
H.~Abdalla\textsuperscript{1}, H.~Abe\textsuperscript{2},
S.~Abe\textsuperscript{2}, A.~Abusleme\textsuperscript{3},
F.~Acero\textsuperscript{4}, A.~Acharyya\textsuperscript{5}, V.~Acín
Portella\textsuperscript{6}, K.~Ackley\textsuperscript{7},
R.~Adam\textsuperscript{8}, C.~Adams\textsuperscript{9}
% ...
\\[1ex]

\noindent
1 : Centre for Space Research, North-West University, Potchefstroom, 2520, South Africa

\noindent
2 : Institute for Cosmic Ray Research, University of Tokyo, 5-1-5, Kashiwa-no-ha, Kashiwa, Chiba 277-8582, Japan

\noindent
3 : Pontificia Universidad Católica de Chile, Av. Libertador Bernardo O'Higgins 340, Santiago, Chile

\noindent
4 : AIM, CEA, CNRS, Université Paris-Saclay, Université Paris Diderot, Sorbonne Paris Cité, CEA Paris-Saclay, IRFU/DAp, Bat 709, Orme des Merisiers, 91191 Gif-sur-Yvette, France

\noindent
5 : Centre for Advanced Instrumentation, Dept. of Physics, Durham University, South Road, Durham DH1 3LE, United Kingdom

\noindent
6 : Port d'Informació Científica, Edifici D, Carrer de l'Albareda, 08193 Bellaterrra (Cerdanyola del Vallès), Spain

\noindent
7 : School of Physics and Astronomy, Monash University, Melbourne, Victoria 3800, Australia

\noindent
8 : Laboratoire Leprince-Ringuet, École Polytechnique (UMR 7638, CNRS/IN2P3, Institut Polytechnique de Paris), 91128 Palaiseau, France

\noindent
9 : Department of Physics, Columbia University, 538 West 120th Street, New York, NY 10027, USA

\noindent
% ...

\end{document}